\documentclass[aps,prfluids]{revtex4-2}
\usepackage{hyperref}
\usepackage{graphicx}
\usepackage{amsmath} 
\usepackage{amssymb}
\usepackage[usenames, dvipsnames]{color}
\usepackage{ifthen}
\newboolean{showcomments}
\setboolean{showcomments}{true}
\usepackage{hyperref}
\usepackage[usenames, dvipsnames]{color}
\usepackage{xifthen}
\newcommand{\dennice}[1]{\ifthenelse{\boolean{showcomments}}
{\textcolor{blue}{Dennice says: #1}}{}}
\usepackage{breqn}%auto line break in equations
\usepackage{fixfoot}%for footnote labels

\DeclareFixedFootnote{\ftAIP}{Reprinted from \citet{RR13}, with permission from AIP Publishing.}
 
\DeclareFixedFootnote{\ftAIAA}{Reprinted from \citet{BMHPLR12} with permission from the authors.} 
 
\DeclareFixedFootnote{\ftAIAAn}{Reprinted from \citet{BMHPLR12} with permission from the authors.} 
 
\DeclareFixedFootnote{\ftEls}{Reprinted from \citet{MSSBU18}, with permission from Elsevier.}

\DeclareFixedFootnote{\ftElsnew}{Reprinted from \citet{MSSBU18}, with permission from Elsevier.}

\DeclareFixedFootnote{\ftElsn}{Reprinted from \citet{MSSBU18}, with permission from Elsevier.}
 
\newcommand{\Rey}{\textit{Re}}
 
\newcommand{\vc}[1]{\ensuremath{\mathbf{#1}}}

\newcommand{\db}{\mathbf{d}}

\newcommand{\psib}{\pmb{\psi}}
\newcommand{\phib}{\pmb{\phi}}
\newcommand{\Phib}{\pmb{\Phi}}

\newcommand{\kk}{\textit{$k^2$}}

\newcommand{\dt}{\partial_t}
\newcommand{\dy}{\partial_y}

\newcommand{\Lap}{\bigtriangleup}

\newcommand{\A}{\mathcal{A}}
\newcommand{\B}{\mathcal{B}}
\newcommand{\G}{\mathcal{G}}

\newcommand{\C}{\mathcal{C}}

\begin{document}

%\preprint{}

\title{Input-output framework for actuated boundary layers} 


%\author{}
%\thanks{}
%\affiliation{}

\author{I.~Gluzman}\email[]{igal.gluzman@gmail.com}\thanks{author to whom correspondence should be addressed}
\author{D.~F.~Gayme}\email[]{dennice@jhu.edu}
\affiliation{Department of Mechanical Engineering$,$ Johns~Hopkins~University$,$ Baltimore$,$ Maryland 21218$,$ USA}


\date{\today}

\begin{abstract}
 This work extends the input-output approach to the study of wall-bounded shear flows manipulated using actuators common in experimental flow control studies. In particular, we adapt this powerful analytical framework to investigate the flow response to specified geometric actuation patterns (e.g., different plasma actuators) that can be applied over a range of different temporal input signals. For example, the commonly studied steady-state (time-averaged) flow response corresponds to a superposition of step responses in our modeling framework. The approach takes advantage of the linearity of the transfer function representation to construct the actuated flow field as a weighted superposition of the flow responses to point sources of varying intensity comprising the actuation model. We first validate the proposed method through comparisons with numerical and experimental studies of the time-averaged behavior of a transitional boundary layer actuated using a dielectric-barrier discharge plasma actuator operating in constricted discharge mode. The method is shown to reproduce the streamwise velocity field and the vortical structures observed downstream from the tested plasma actuator configurations. We then demonstrate that the method provides even better agreement with the steady-state response of the boundary layer subject to actuation from arrays of symmetric plasma actuators arranged in both spanwise and serpentine geometries. These results indicate the utility of this extension to the widely used input-output framework in analyzing the effects of certain actuation modalities that have shown promise in flow manipulation strategies for drag reduction. An important benefit of this analytical method is the low computational cost associated with its use in extensive parametric studies that would be cost-prohibitive using experiments or high-fidelity simulations.
\end{abstract}

%\keywords{}

\maketitle


\section{Introduction}

Flow modification through a range of control actions, such as surface blowing and suction \citep{MKSK06}, the introduction of transverse wall oscillations \citep{MJ12}, vortex generators \citep{HLNM10}, and constrictive discharge plasma actuation \citep{MSSBU18} has shown success in producing desired changes to the flow characteristics. However, many of these approaches are known to produce the desired behaviors over only limited parameter ranges. For example, blowing and suction induced traveling waves reduce drag over a limited range of wave speeds and amplitudes, and a single propagation direction \citep{MKSK06,MJ10}. A vortex-generator array is effective in mitigating the transition for a limited range of spanwise spacings between array elements \citep{HLNM10}. Similarly, the frequency, duty cycle, and other input signal properties have been shown to play a role in the efficacy of these and other types of actuators used in flow modification \citep{CS11}. Understanding the potential of different flow control strategies requires a full characterization of the parameter range and input signals that produce the desired response. However, obtaining such knowledge through experimental or numerical studies can become cost-prohibitive when the range of conditions that need to be tested is large. More efficient use of these approaches can be achieved by first employing analysis techniques that enable a qualitative understanding of the effect of different actuation signals on the flow fields \citep{KB07}. 

Input-output-analysis-based tools have shown great promise in providing the required understanding. They have been widely used to provide insight into the flow characteristics arising from structured external forcing, e.g., stochastic forcings \citep{FI93,BD01}, impulsive forcings \citep{JB01,JB05} and harmonic forcings \citep{MS10}. In particular, the externally forced linearized Navier-Stokes system, has shown success in examining the important dynamic processes, structural features and energy pathways of transitional \citep[e.g.,][]{FI93,BD01,JB01,JB05} and turbulent \citep[e.g.,][]{AJ06,HC10} wall-bounded shear flows. 
 
 In flow control applications, input-output-based analysis has been used to derive control laws \citep{BHHS09,SBBH13} that were successfully applied in laminar boundary layers in experiments and used for preliminary assessment of actuation strategies \cite{FSFGBH15,TVK19}. These methods have also been adapted to analyze the effect of flow manipulation through streamwise traveling waves generated by surface blowing and suction \citep{MJ10}, as well as to design transverse wall oscillations that suppress turbulence in a channel flow \citep{MJ12}. Optimal riblet shapes for drag reduction in turbulent channel flow have also been obtained through resolvent analysis~\citep{CL19}. These and a host of other works demonstrate the promise of an input-output based approach in identifying promising flow manipulation strategies. However, periodic actuation such as transverse wall oscillations and surface blowing and suction can pose implementation challenges~\cite{Ga07}. 
 
On the other hand, plasma actuators are easier to implement and are becoming increasingly common in experimental flow control studies \citep{CPO09}. Plasma actuators come in a vast number of configurations and can be operated under pulsed excitation with adjustable duty-cycle and frequency~\citep{KGVS11,MC11,RR13}. A number of these actuation strategies have shown promise in separation control~\citep{YK17}, attenuation of Tollmien–Schlichting waves \cite{BTG15}, turbulent drag reduction \citep{TCDMY19}, and control of shock-waves~\cite{KSM18}. The operation of plasma actuators relies on discharge between the electrodes of the actuator, which leads to a localized input that is typically modeled as a body force distribution in physical space~\citep{RR13,SKH18, MSHCH20}. The need to capture a geometric distribution of forcing in physical space complicates direct application of the input-output paradigm, which typically relies on transforming the problem to streamwise and spanwise wave number space and then obtaining the solution to problems in terms of a superposition of the effects at each streamwise and spanwise wavenumber pair. This work provides a means to adapt the framework to address this limitation.

The main contribution of this work is a method to extend the input-output modeling paradigm to compute the flow response to specific localized actuator geometries with input signals that can be represented through a parametrized family of pulse-width modulated excitations. This input signal can be adapted to model a variety of input signals, including steps, impulses, and periodic and short pulses. 
Previous works have demonstrated that important knowledge of the flow response is possible through models that capture the local effect of an actuator on the flow field \citep{BHHS09,SBBH13,FBH17}. We adopt this idea and develop a model of the effect of particular actuator geometries on the flow field as a geometric arrangement of point source inputs of varying intensity and input directions. The output associated with the pulse modulated signal can then be computed analytically for each point source. We then exploit the linearity of the input-output technique to construct the flow response as a superposition of the weighted point source response functions. We validate the method for the special case of step input signals, which provide a model for continuous actuation signals. This input class provides an important test case as the associated flow response corresponds to the time-averaged flow field due to constant actuation, which is extensively reported in the literature. 
 
We test our model against the results from both numerical and experimental studies of three different classes of plasma actuators that are used to reduce transient growth in a Blasius boundary layer. We first demonstrate that the approach reproduces the vortical structures, the high and low-speed streaks, and streak spacing associated with the vortices generated due to a single exposed electrode dielectric-barrier discharge (DBD) plasma actuator operated in constricted discharge mode \citep{MSSBU18}. We then focus on two geometric patterns; a linear spanwise array of symmetric DBD plasma actuators \citep{BMHPLR12} and a serpentine DBD plasma actuator \citep{RR13}. In both cases, the model reproduces the structural features of the time-averaged velocity fields obtained from the validation data. Our results indicate that the proposed analytical approach can evaluate the efficacy of these common streak and vortex generation mechanisms, used to reduce drag and control transition, mixing, and separation in boundary layers \citep{JR98,FT12}.

The remainder of the paper is organized as follows; the model derivation is presented in Sec.~\ref{sec:iomodel}. Validation of the model for the three different actuation geometries is provided in Sec.~\ref{sec:results}. Finally, concluding remarks and future directions are discussed in Sec.~\ref{sec:confuture}.

\section{Analytical model of actuated boundary layers}
\label{sec:iomodel}

We consider incompressible wall-bounded parallel shear flow with streamwise direction $x$, wall-normal direction $y$, and spanwise direction $z$. We decompose the velocity field into a base flow of the form $\vc{U}=\begin{bmatrix} U(y) & 0 & 0 \end{bmatrix}^T$ and perturbations about that base flow $\vc{u}=\begin{bmatrix} u & v & w\end{bmatrix}^T$. 

We compute the effect of actuation on the flow field as a solution of the linearized Navier-Stokes equations about the base flow, $\vc{U}$, subject to body forcing. Spatial invariance of the parallel flow field in the horizontal directions enables us to evaluate these equations through their $(x,z)$ spatial Fourier transform 
\begin{equation}
 \dt\psib(k_x,y,k_z,t) = \A(k_x,y,k_z) \psib(k_x,y,k_z,t) + \B(k_x,y,k_z) \db(k_x,y,k_z,t).
 \label{eq:LNSkxkz}%
\end{equation}%
Here $\psib:=\begin{bmatrix} \hat{v} & \hat{\omega}_y \end{bmatrix} ^T$ is the state vector comprised of the transformed wall-normal velocity $\hat{v}$, and wall-normal vorticity $\hat{\omega}_y$ parametrized by the respective streamwise and spanwise wave-numbers, $k_x$, and $k_z$. The vector $\db(k_x,y,k_z,t)=\begin{bmatrix} \hat{d}_x & \hat{d}_y & \hat{d}_z\end{bmatrix}^T$ describes the transformed body forcing. The operator 
\begin{equation}
\A :=
		\begin{bmatrix}
		-i k_x \Lap^{-1} U \Lap+i k_x\Lap^{-1}U''+(\Rey \Lap)^{-1}\Lap^2 & 0 \\ 
		-i k_z U' & -i k_xU+\Rey^{-1}\Lap
		\end{bmatrix}, 
\end{equation}
where $U':=dU(y)/dy$, $\Lap:=\partial_{yy}-\kk$, $\kk:=(k_x^2+k_z^2)$, and 
\begin{equation}
\B :=
\begin{bmatrix}
		\B_x & \B_y & \B_z
		\end{bmatrix}
:=
			\begin{bmatrix}
			-i k_x\Lap^{-1}\dy \gamma & -\kk\Lap^{-1} \gamma& -i k_z\Lap^{-1}\dy \gamma\\ 
			 i k_z \gamma& 0 & -i k_x \gamma
			\end{bmatrix} \label{eq:B}	
\end{equation}
shapes the forcing through the function $\gamma(y)$ \citep{JB05}. 

The response of the velocity field to the given input is described through the output equation
\begin{equation}
 \phib(k_x,y,k_z,t) = \C(k_x,y,k_z) \psib(k_x,y,k_z,t),
 \label{eq:LNS_output}
\end{equation}
where
\begin{align}
\C &:=
\begin{bmatrix}
		\C_u\\
		\C_v\\
		\C_w
		\end{bmatrix}
:=
 		\frac{1}{\kk}
		\begin{bmatrix}
		i k_x\dy & -i k_z \\
		\kk & 0 \\
		i k_z\dy 	& i k_x
		\end{bmatrix} \label{eq:C}
\end{align}
transforms the state vector $\psib$ into the output vector $\phib=\vc{\hat{u}}=\begin{bmatrix} \hat{u} & \hat{v} & \hat{w} \end{bmatrix}^T$.

Our model for different actuation geometries is built upon spatially localized forcing at a specific wall-normal location, $y_0$. We adopt the approach in \cite {JB01} and represent the forcing at this location as a normal distribution with mean $y_0$ and variance $2\epsilon $, i.e.,
\begin{equation}
\gamma(y)=\frac{1}{2\sqrt{\pi \epsilon}}e^{-\frac{(y-y_0)^2}{4\epsilon}}, \:\: \epsilon>0, 
\label{eq:forcing} 
\end{equation}
where $\epsilon$ is sufficiently narrow to ensure that the forcing is concentrated at $y_0$. We consider actuation signals that can be represented as special cases of pulse train body forcings of the form 
\begin{equation}
d(t)=\sum\limits_{n=0}^{N-1} {\left[H(t-nT) -H(t - nT - \tau)\right]},
\label{eq:pulse}
\end{equation}
where $H(t)$ is a unit step function, $N$ is the number of pulses, $T$ is the period between pulses, and $\tau\in{(0,T)}$. The parameters $\tau$, $T$, and $N$ in Eq.~\eqref{eq:pulse} can be adjusted to represent a number of temporal signals, e.g. an impulse (with sufficiently short $\tau$ and $N=1$) or an impulse train with $N>1$. A step input can be obtained by setting $N=1$ and $\tau>t$.   
The response to inputs of the form~\eqref{eq:pulse} can be computed as \citep{He18} 
\begin{equation}
\phib(t)=\C\A^{-1}\!\!\sum\limits_{n=0}^{N-1}\! {\left[ \left( e^{\A\left( {t - nT} \right)} - I \right)\!{H(t-nT)} -\! \left( e^{\A\left(t - nT - \tau \right)} - I \right)\!\! H(t - nT - \tau) \right]}\G, 
\label{eq:sol_pulse}
\end{equation}%
where $\G:=\B_x+\B_y+\B_z$. 

We next exploit the linearity of the system dynamics in Eq.~\eqref{eq:LNSkxkz} and Eq.~\eqref{eq:LNS_output} to develop a method to compute the flow response to actuation over a spatial pattern. In particular, we build the desired response function as a superposition of the velocity fields due to weighted point source inputs arranged in a pattern that models the desired flow actuation. This extension of well-known input-output techniques, see e.g. \citep{JB01,Jo04,HJK18}, enables its application to common experimental actuator configurations that may not be well represented as a single point input or more general body forcing, e.g. delta-correlated stochastic forcing \citep{FI93,JB05}. 

\begin{figure} 
\centering
\includegraphics[trim={0 0 0 0},clip,width=1\linewidth]{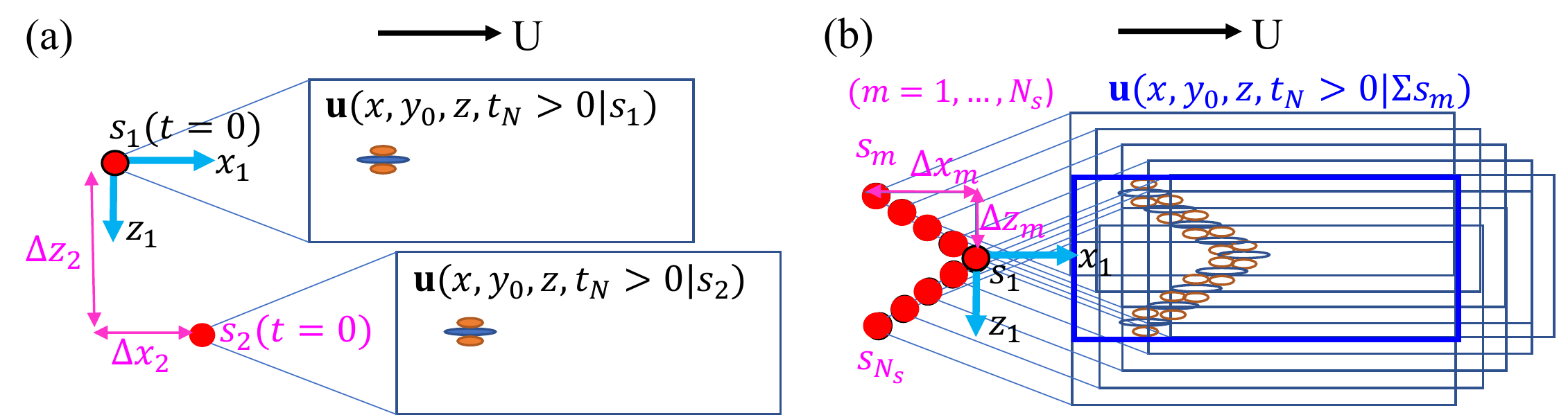}
	\caption{Conceptual sketches of the streamwise velocity fields ($xz$-plane at the wall-normal location $y_0$) at time $t>0$ due to individual point sources indicated by red circles. (a) Responses $u(x,y_0,z,t|s_1)$ and $u(x,y_0,z,t|s_2)$ due to respective sources $s_1$ and $s_2$. (b) The bold blue rectangle represents the streamwise velocity field $u(x,y_0,z,t|\sum s_m)$ due to an actuator modeled as a superposition of $N_s$ sources organized in a triangular pattern. 
	\label{fig:scheme}}
\end{figure}

Consider a single spatially localized input (source), denoted as $s_1$, at horizontal location $(x_1,z_1)$, which we assign as the origin. The location of a second source $s_2$ can then be described in terms of distances $\Delta x_2=x_2-x_1$ and $\Delta z_2=z_2-z_1$ from this origin, as shown in Fig.~\ref{fig:scheme}a. The $(x,z)$ spatial Fourier transform of the flow field arising from source $s_2$ can then be computed as
\begin{equation}
\phib(k_x,y,k_z,t|s_2) = e^{ - i(k_x\Delta x_2+k_z\Delta z_2)}\phib(k_x,y,k_z,t|s_1),
\label{eq:s_m}
\end{equation}
where $e^{ - i(k_x\Delta x_2+k_z\Delta z_2)}$ results from a shift theorem \citep{Sm07} and $\phib(k_x,y,k_z,t|s_1)$ represents the transformed flow response to a point source at the origin $(x_1,y_1)$. 

The linearity of the model in Eq.~\eqref{eq:LNSkxkz} and the expression in Eq.~\eqref{eq:s_m} enables the construction of arbitrary geometric actuation patterns as an array of weighted $N_s$ sources, with each source $s_m$ is shifted by $\Delta x_m$ and $\Delta z_m$ from the predefined origin $(x_1,z_1$), as illustrated in Fig.~\ref{fig:scheme}b. The flow response due to this patterned actuation can then be computed as
\begin{dmath}
\Phib(k_x,y,k_z,t)=\!\frac{1}{N_s}\!\sum_{m=1}^{N_s}{ \!\!\left[ e^{ - i(k_x\Delta x_m+k_z\Delta z_m)}c_m\!\!\!\!\!\sum_{j=x,y,z}\!\!\!\!{e_{d,j}(m)\phib_{j}(k_x,y,k_z,t|s_m)}\right]},
\label{eq:sum_kxkzeff}
\end{dmath} 
where $c_m(k_x, y,k_z)$ is a weighting function that takes values ${[-1,1]}$ and assigns a relative amplitude to each source $s_m$ with respect to source $s_1$. Negative values of $c_m$ represent forcing in the negative direction. The normalization $N_s$ in Eq.~\eqref{eq:sum_kxkzeff} ensures that the response is invariant to the number of sources. Forcing in a given direction is defined through the term $\sum_{j=x,y,z}{e_{d,j}(m)\phib_{j}(k_x,y,k_z,t|s_m)}$, which is a weighted sum of the responses of source $s_m$ to forcing in an axial $(x, y, z)$ direction $j$. The weights are defined through the unit vector $e_{d}=\begin{bmatrix} e_{d,x} & e_{d,y} & e_{d,z}\end{bmatrix}^T$ and $\phib_{j}(k_x,y,k_z,t|s_m)$ is obtained by setting $\G = \B_j$ in Eq.~\eqref{eq:sol_pulse}, where $\B_j$ for each direction $(x, y, z)$ is defined in Eq.~\eqref{eq:B}.

In the next section, we employ the analytical approach described above to compute the response of a transitional boundary layer to a spanwise array of symmetric DBD plasma actuators, and a DBD plasma actuator operating in constricted~discharge~mode. We focus on the steady-state step response, which corresponds to the time-averaged flow fields arising due to continuous actuation. The response to a point source input of this form applied in the $j$ direction can be computed as a special case of Eq.~\eqref{eq:sol_pulse} in the $\lim_{t\rightarrow\infty}$ with $t<\tau$, and $N=1$, which leads to functions of the form 
\begin{equation}
\phib_j(k_x,y,k_z) =-\C \A^{-1} \B_j.
\label{eq:step_ss} 
\end{equation}
 We note that although the numerical results in this work focus on a single type of actuation signal, the procedure described above can be used also to study the flow response due to an impulse and or an impulse train applied in the $j$ direction by replacing the function $\phib_j$ in Eq.~\eqref{eq:sum_kxkzeff} with 
$\phib_j(k_x,y,k_z,t)=\C e^{\A t} \B_j$ or $\phib_j(k_x,y,k_z,t)=\C \sum_{n=0}^{N-1}e^{\A (t-nT)}\B_j$, respectively.

\section{Results}
\label{sec:results}
 
We now demonstrate the efficacy of the approach described in Sec.~\ref{sec:iomodel} in reproducing the steady-state response of the  
transitional boundary layer to a continuous input from three different types of DBD plasma actuation: a DBD actuator operating in constricted discharge mode (Sec.~\ref{sec:constircted}) and arrays of DBD actuators arranged both in a line along the spanwise direction (Sec.~\ref{sec:DBDVG}) and in a serpentine geometry (Sec.~\ref{sec:serpantince}). For each case we first detail the actuation model in terms of its effect on the flow and then describe the corresponding response in both the velocity and vorticity field based on Eq.~\eqref{eq:sum_kxkzeff} with the appropriately defined output matrix in Eq.~\eqref{eq:C}.

 We compute the flow response by discretizing the operators in the wall-normal direction using Chebyshev co-location points, which leads to a Fourier-Chebyshev-Fourier discretization of the problem. We employ a quasi-parallel assumption and use the change of variables described in \citet{SH01} to transform the bounded domain $\begin{bmatrix} -1, & 1 \end{bmatrix}$ to the semi-infinite domain $\begin{bmatrix} 0, & \infty \end{bmatrix}$ of the flat plate boundary layer (for details see Appendix A.4 in \citep{SH01}). We set the wall-normal domain range to $[0, L_y]=\begin{bmatrix} 0, & 15 \end{bmatrix}$ as this height is well within the free stream and thus captures the entire base flow profile variation. The computations are performed in MATLAB$^\circledR$ \emph{R2017B} with differentiation in the wall-normal direction implemented via the pseudo-spectral differentiation matrices of \cite{WR00}. We employ $N=40$ Chebyshev grid points over the pre-transformation wall-normal extent of $\begin{bmatrix} -1, & 1 \end{bmatrix}$, and use $2048\times 256$ linearly spaced grid points over $\{k_{x,min} := -5, k_{x,max} :=4.98\}$ $\times$ $\{k_{z,min} := -18, k_{z,max} := 17.86375\}$. We verify that the selected $N$ and the $(k_x,k_z)$ range are sufficient by doubling the domain and the number of points and verifying that the changes to the observed structures are negligible. 

In all cases, we employ a Blasius base profile in Eq.~\eqref{eq:LNSkxkz}. The Reynolds number is determined through nondimensionalizeation of the flow parameters by the displacement thickness $\delta^*$ and the free stream velocity $U_\infty$. 
The forcing is applied at the grid point closest to the wall (to simulate forcing at the wall). This location corresponds to $y_0=0.02$ in Eq.~\eqref{eq:forcing} for the Chebyshev grid with $N=40$. For each point source used to construct the representation of the actuation input we specify the width of the forcing function as $\epsilon= 5\times 10^{-4}$. This choice of $\epsilon$ is validated by decreasing the value to $5\times 10^{-7}$, i.e., increasing the intensity and decreasing the width of the forcing, and then verifying that the effect on the flow structures is negligible. The choice is also consistent with the value used by \citet{Jo04} and \citet{HJK18}, who studied the response to localized body forces in channel flows. We build the actuator model as a superposition of these weighted point sources.

\subsection{Plasma actuator operating in constricted discharge mode} 
\label{sec:constircted}

\begin{figure}
\centering
	\includegraphics[trim={0 0 0 0},clip,width=1\linewidth]{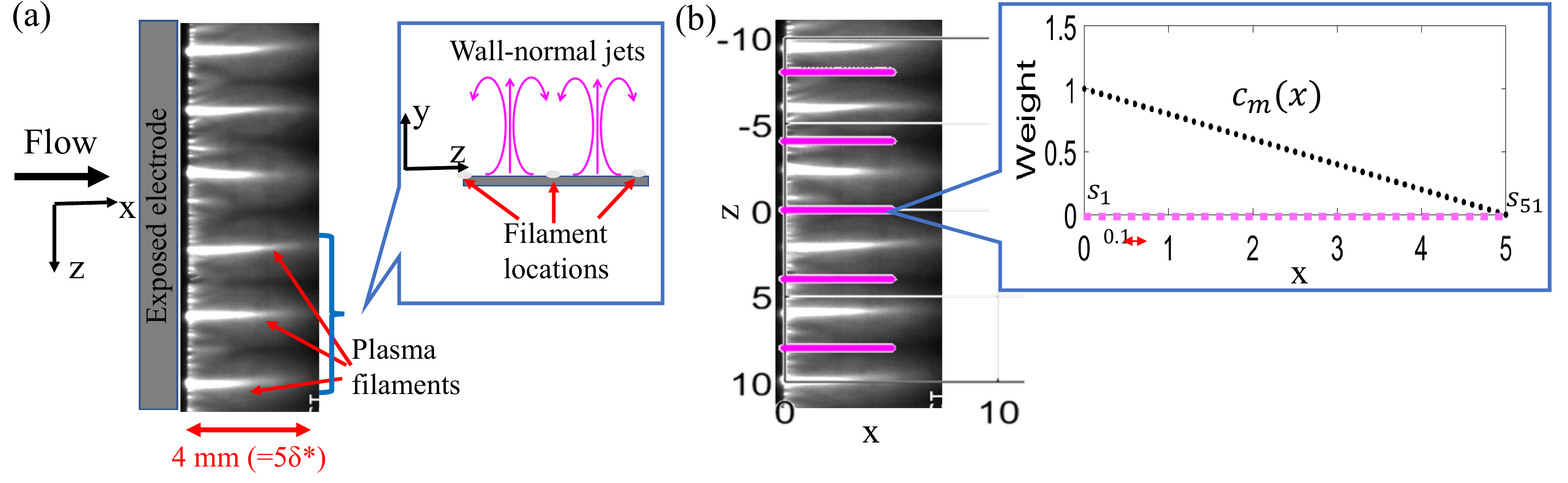} 
\caption{(a) The $xz$-plane view of the actuator operating in constricted discharge mode. We denote the exposed electrode by a gray rectangle. The constricted filaments (plasma bursts) with approximate spacings of 2.5~mm are indicated in bright white. This cropped image of the discharge is taken from \citet[figure 4]{MSSBU18}\ftEls. (b) Schematic of our actuator model comprised of an array of sources in the form of streamwise ribs placed between filament pairs. Each array consists of 51 sources equally spaced at a distance of 0.1 in dimensionless units (nondimensionalized by $\delta^*$). In the figure, each source is denoted by a magenta dot as noted in the blow up of the forcing on the far right. A steady-state step response to wall-normal forcing is used and weighted as $\tilde {c}_m(x)=1-0.2x$ (in physical space and dimensionless units, nondimensionalized by $\delta^*$), which corresponds to the variation in filament intensity as a function of streamwise distance.
\label{fig: xzc}}
\end{figure}

In this section, we model a DBD plasma actuator operating in a constricted discharge mode configuration based on the setup in \cite{MBKB14, MSSBU18}. We then compare the steady-state flow response computed from Eq.~\eqref{eq:sum_kxkzeff} with Eq.~\eqref{eq:step_ss} to that obtained in the experiments in \cite{MSSBU18}. The actuator configuration consists of a single exposed electrode whose upstream edge is located $200$~mm downstream from the geometric leading edge of the plate. The measured displacement thickness at the actuator is $\delta^*$=0.81~mm, the freestream velocity is $U_\infty=12$~m/s and the Reynolds number is $\Rey=U_\infty\delta^*/\nu=650$.

Figure~\ref{fig: xzc}a illustrates how the operation of the DBD actuator in constricted discharge mode affects the flow field at the actuation site (see also Fig. 4 in \citep{MSSBU18}). The actuation introduces plasma filaments that are elongated in the streamwise direction (the bright regions in Fig.~\ref{fig: xzc}a). These produce wall-normal fluid jets between the filaments (illustrated in the inset in Fig.~\ref{fig: xzc}a) that decay with streamwise distance along the filament. It is this vertical injection of velocity in the flow that we model, as it is directly responsible for the induced vorticity that is the goal of the constricted discharge actuation. 

We model the vertical injection of fluid as an array of sources centered between the plasma filaments organized in the form of streamwise ribs at the plate surface. This configuration is shown in Fig.~\ref{fig: xzc}b, where the locations of the sources are denoted by magenta dots. We set the dimensionless rib length to $5$, which corresponds to our computed value for the average length of the bright regions in the experimental data, as indicated in the annotation in Fig.~\ref{fig: xzc}a. We model each of the ribs as a cluster of 51 sources spaced at 0.1 dimensionless units apart where the intensity of the forcing in the downstream direction decays as $1-0.2x$. The selected slope of $-0.2$ is determined through a linear fitting of the intensity decay of a single representative filament. Our model corresponds to a uniform spacing between the filaments of $3.7\delta^*$, based on the average filament spacing in Fig.~\ref{fig: xzc}a. However, we note that the experiment has an inhomogeneous distribution of the discharge's plasma density (in a classical electrodes arrangement used in \cite{MBKB14} for that plot), which results in nonuniform spacing between the filaments. We then compute the flow response to continuous actuation by applying Eq.~\eqref{eq:sum_kxkzeff} with weighting functions of the form $\tilde{c}_m(x)=1-0.2x$ (in physical space) and $\phib_j(k_x,y,k_z)$ from Eq.~\eqref{eq:step_ss} with $j=y$.

\begin{figure}[b]
\centering
\includegraphics[trim={0 0 0 0},clip,width=0.8\linewidth]{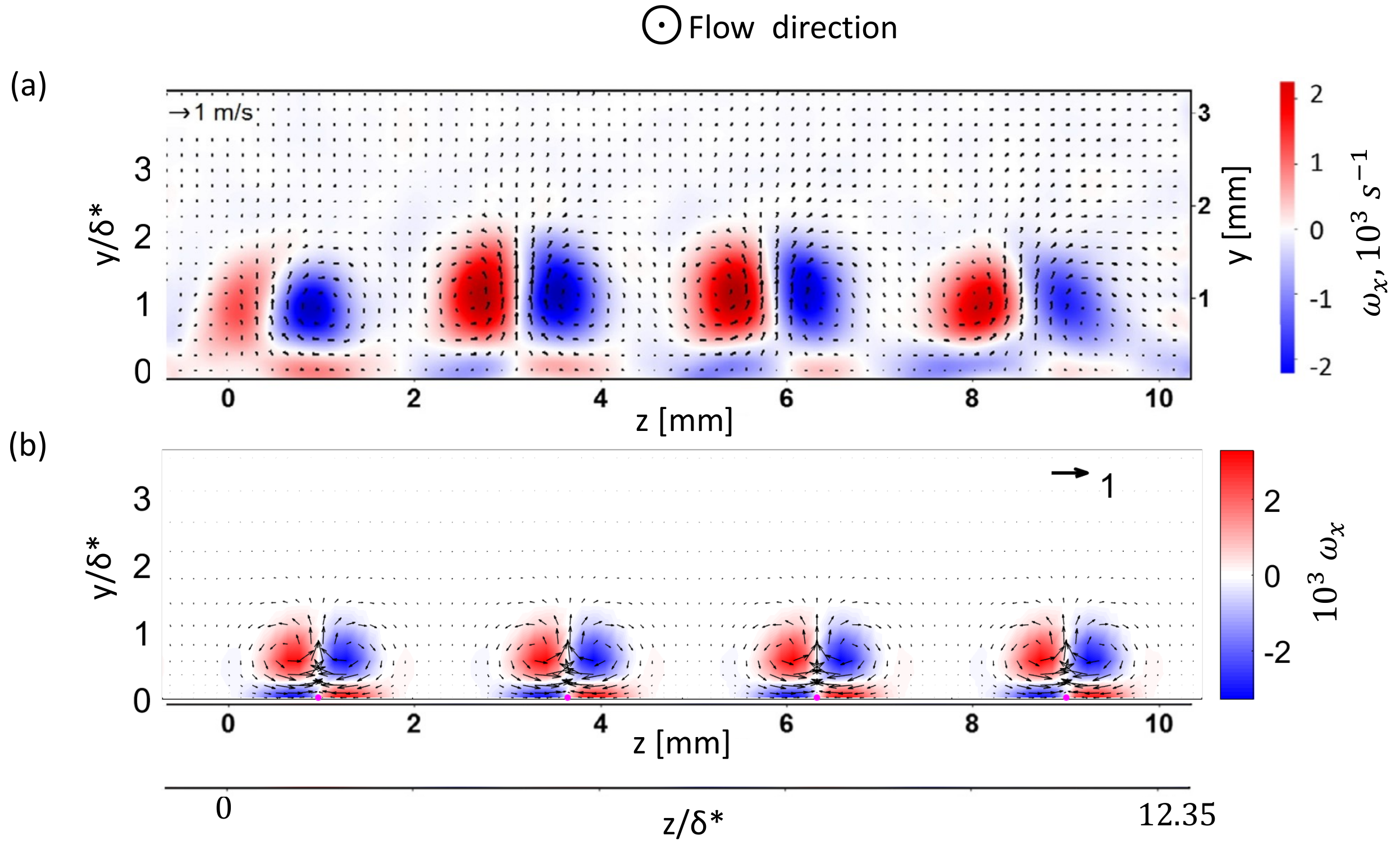}
\caption{Crossflow plane showing the streamwise vortical structures ($\omega_x=\partial_y w- \partial_z v$) of the actuated boundary layer. (a) PIV measurements in Fig.~10a in \citep{MSSBU18}\ftElsnew at $x/\delta^*=12.6$ (corresponds to 10~mm). (b) Our model at {$x/\delta^*=12.6$} for $u_{max}/U=0.25$ and a rib gap of $3.7\delta^*$ (3~mm). 
\label{fig: vortx_yzc}}
\end{figure}

 The longitudinal vortex pairs arising due to actuation in the $yz$-plane at 10~mm ($12.6\delta^*$) downstream from the actuator are shown in Fig.~\ref{fig: vortx_yzc}.  Figure~\ref{fig: vortx_yzc}a provides the vortical fields computed from the experimental data of \citep[Fig. 10a in][]{MSSBU18}~and Fig.~\ref{fig: vortx_yzc}b provides results from the proposed approach. 
 A comparison of the two fields indicates good qualitative agreement in terms of the structural features. In particular, the four counter-rotating vortices between each filament pair, which are the most significant features of the actuated flow, are reproduced. However, the vortical structures arising from our analysis are more localized, which is likely related to the simplified actuation model that only perturbs the flow between the filaments. We note that the effect of the nonuniform spacing in the experiment affects both the shape and the spacing of the structures, with some being closer in aspect ratio to those obtained from the model than others. It is unclear if the differences observed arise from the simplified actuation model that only applies forces in the wall-normal direction or if they are due to nonlinear interactions that are not captured in the framework. Isolating these effects would require the development of a detailed actuator model and the introduction of nonlinear effects, which are both beyond the scope of this study. Refinement of the model representing this type of actuation to better understand this discrepancy is a direction for future work.

 The qualitative agreement observed in this example motivates the next two sections, which investigate whether the approach is well suited to arrays of DBD actuators. These configurations are also commonly used in experiments yet their affect on the flow are more straightforward to represent in the proposed framework.

\subsection{Spanwise array of symmetric DBD plasma actuators} 
\label{sec:DBDVG}

\begin{figure}
\centering
	\includegraphics[trim={0 0 0 0},clip,width=1\linewidth]{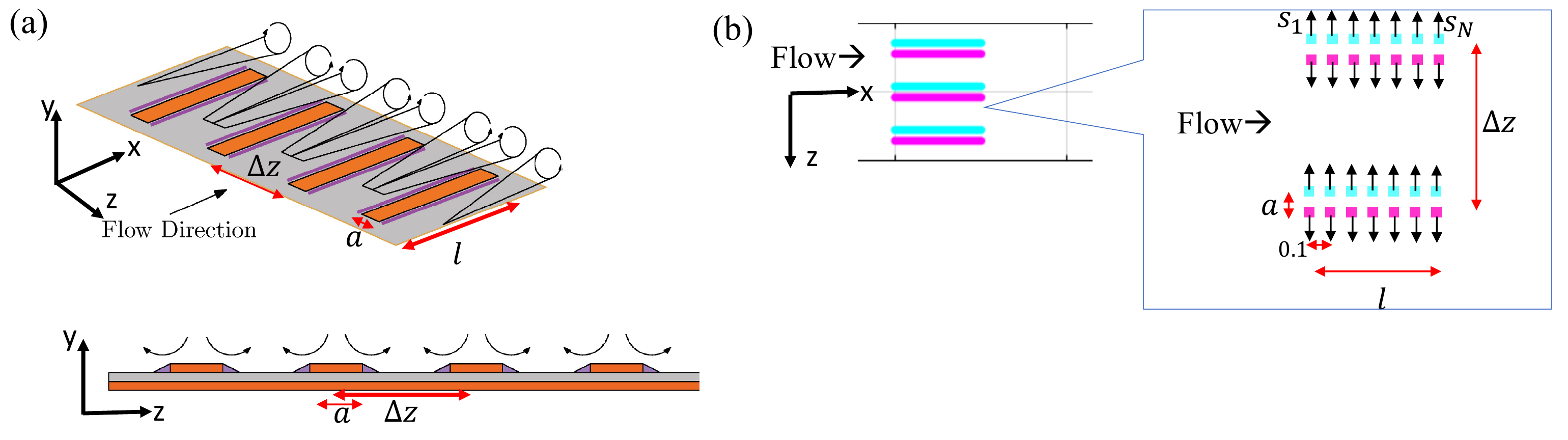}
	\caption{(a) Experimental and computational setup with a schematic view of the plasma actuator array (taken from \citet[figure 3]{BMHPLR12}\ftAIAA). (b) Each actuator is modeled as two linear arrays of forcing. The lower line (shown in magenta) applies forcing in the positive spanwise direction and the upper line (shown in cyan) applies forcing in the negative spanwise directions. Together these forcings induce the sideways motion indicated in the lower schematic of (a). 
\label{fig:setupAIAA}}
\end{figure}

We model the spanwise array of symmetric DBD plasma actuators in the experimental configuration of \citet{HLNM10} and associated  direct numerical simulations  (DNS) of \cite{BMHPLR12}. We then compare the steady-state flow response computed from Eq.~\eqref{eq:sum_kxkzeff} with Eq.~\eqref{eq:step_ss} to that obtained in \citet{HLNM10} and \cite{BMHPLR12}. Figure~\ref{fig:setupAIAA}a provides the schematic from Fig.~3 in \citep{BMHPLR12} 
%\footnote{\label{AIAA}{ Reprinted from \citet[figure 3]{BMHPLR12} with permission from the authors.}} 
 of the actuator location and geometry. Here, the plasma actuator array is located 250~mm downstream from the geometric leading edge of the plate and extends $l=$40~mm in the streamwise ($x$) direction. The width of each exposed electrode is $a=5$~mm and the electrodes are spaced $\Delta z=20$~mm apart. The displacement thickness at the upstream edge of the actuator is $\delta^*$=1.59 mm, and the Reynolds number is $\Rey=U\delta^*/\nu=530$, based on a freestream velocity of $U_\infty=5$~m/s. 

We model each electrode in the spanwise array of DBD actuators as two lines of point sources, indicated by magenta and cyan lines in Fig.~\ref{fig:setupAIAA}a. We apply an outward forcing to each point in the cluster, which corresponds to forcing in the positive spanwise direction for the points in magenta in Fig.~\ref{fig:setupAIAA}a, and the negative spanwise direction for the points in cyan. We impose a streamwise spacing between the sources of 0.1 in dimensionless units, which results in 252 sources along each exposed electrode edge to model the full $40$~mm length of the actuator. We build the array of four actuators spaced 12.6 nondimensional units ($20$~mm) apart. The response is computed using Eq.~\eqref{eq:sum_kxkzeff} with the steady-state step response defined in Eq.~\eqref{eq:step_ss} with $j=z$. Sources that apply forcing in the positive $z$ direction are assigned weights of $c_m=1$, whereas a weighting of $c_m=-1$ is assigned to sources that apply forcing in the opposing direction.

\begin{figure}
\includegraphics[trim={0 0 0 0},clip,width=1\linewidth]{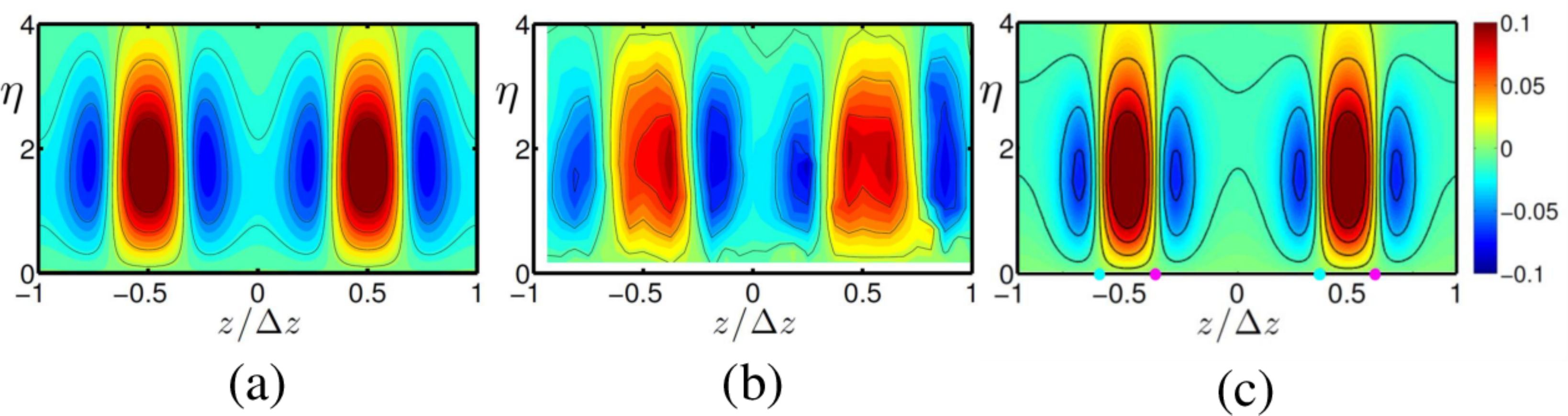} 
\caption{
Plots of $yz$-plane contours of the normalized streamwise component of the perturbation velocity ($u/U_\infty$) at 200~mm ($x/\delta^*= 125.8$) downstream from the actuator from (a) DNS, (b) experiments taken from \citet[figure 11]{BMHPLR12}\ftAIAAn, and (c) the present model. Here $\eta= y\sqrt{(U_\infty/\nu x_0)}$ is the Blasius length scale, where $x_0$ is the streamwise distance from the leading edge of the plate. The electrode edges are indicated with magenta and cyan points. The results in (c) are scaled as $u_{max}/U=0.4$ to match the contour color range of the experimental results. \label{fig:uyzsimexpmodel}} 
\end{figure}

Figure~\ref{fig:uyzsimexpmodel} shows contours of the normalized streamwise component of the perturbation velocity ($u/U_{\infty}$) at a distance of 200~mm ($x/\delta^*= 125.8$) downstream from the electrode array obtained through DNS \citep{BMHPLR12} (panel a) and experiments \cite{HLNM10} (panel b) from \citet[figure 11]{BMHPLR12}.
%\footnote{Reprinted from \citet[figure 11]{BMHPLR12}, with permission from the authors.}.
 Figure \ref{fig:uyzsimexpmodel}c provides results from the proposed approach, where the output is scaled as $u_{max}/U=0.4$ to match the contour color range of the experimental results. The DNS data were also scaled to match the experimental results (see \citep{{BMHPLR12}} for details). The plots demonstrate that our model obtains good qualitative agreement in terms of the shape of the flow structures with both the DNS and experiments. The streaks of streamwise velocity show quantitative agreement with the spanwise spacing of the actuator electrodes in the array with low-momentum regions between the electrode pairs. 

Having validated the model's performance for DBD actuators arranged in a simple geometric pattern, we next examine its performance in the more complex serpentine pattern.

 \subsection{The serpentine geometry plasma actuator} 
\label{sec:serpantince}

\begin{figure}
\centering
	\includegraphics[trim={0 0 0 0},clip,width=0.8\linewidth]{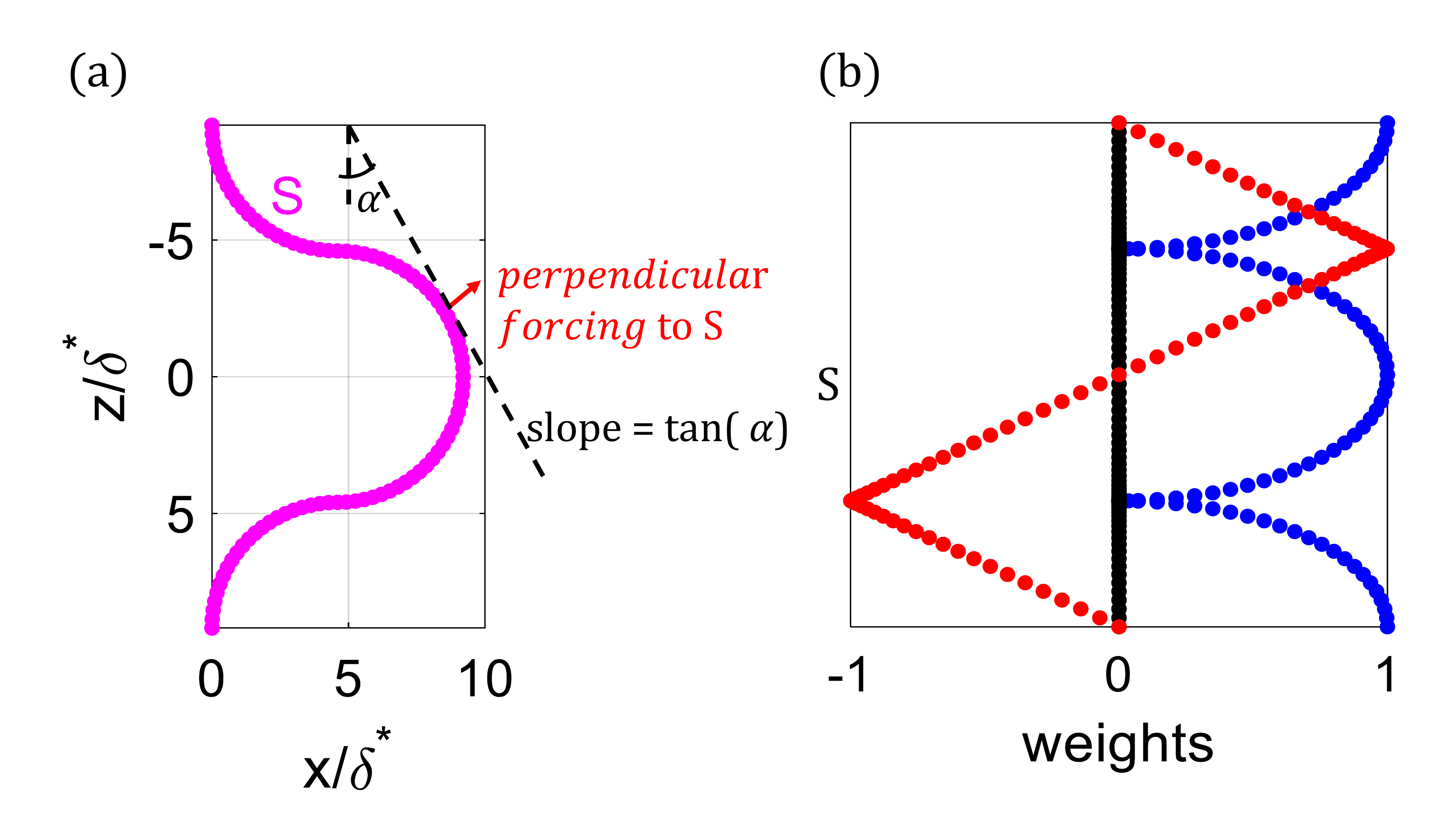} 
\caption{(a) Reproduced\ftAIP $xz$-plane view of the serpentine actuator used in \citet[figure 2c]{RR13}. A cluster of 82 equally spaced point sources (41 sources for each semi-circular segment) denoted by magenta dots form one wavelength of the circular pattern. The radius of the semi-circular arcs is 7.5~mm in our study and the experiment. A steady-state step response to perpendicular forcing defined through the weighted sum of the responses of each to forcing in an axial direction $(x, y, z)$ is studied. (b) Weights for each axial direction as a function of the arc length $S$: $e_{d,x}$ (blue),$e_{d,y}$ (black),and $e_{d,z}$ (red).
\label{fig: xzc_serp}}
\end{figure}

We model the steady-state flow response to actuation from the DBD serpentine plasma actuator described in \cite{DR12,RR13} and compare the predictions from the proposed approach, with Eq.~\eqref{eq:step_ss}, to simulation data from \citet{RR13}. The setup is shown in Fig.~\ref{fig: xzc_serp}a, which provides a schematic from \citet[figure 2c]{RR13} depicting one segment (a single wavelength) of the actuator geometry. Here, the serpentine geometry plasma actuator is centered at $x/L=1.025$, where $L=300~mm$ is the downstream distance from the computational box inlet. The geometry consists of patterned circular arcs with a radius of $7.5$~mm. The displacement thickness at the actuator's upstream edge is $\delta^* =1.63$~mm, and the Reynolds number is $\Rey=U_\infty\delta^*\nu=544$, based on a freestream velocity of $U_\infty=5$~m/s. Further details regarding the geometry and setup are provided in \cite{RR13}. 

In the simulation of \cite{RR13}, the serpentine plasma actuation is modeled as a steady application of outward body force perpendicular to the geometry edge. The simulation was validated against an experimental study \cite{DR12}, which examined the effect of the actuation under quiescent conditions. In both studies it was shown that a vectored jet is produced at the pinch points of the actuation (the upstream concave part of the geometry), whereas a simple wall jet forms at the spreading point (the downstream convex part of the geometry). It is these effects that we seek to reproduce in our model of the actuation.

We represent the outward body forcing perpendicular as a superposition of weighted responses in the streamwise and spanwise directions. The shapes of the weighting functions $e_{d,x}=cos(\alpha)$ and $e_{d,z}=sin(\alpha)$ are shown in Fig.~\ref{fig: xzc}b, where $\alpha$ is a slope angle of the contour at each source location (see Fig.~\ref{fig: xzc}a). This forcing is applied as 41 point sources spaced along each semi-circular arc of radius $4.66\delta^*$ ($7.5$~mm), which corresponds to a spacing of 0.0766 ($\pi/41$) rad. These points are shown as magenta dots in Fig.~\ref{fig: xzc}a. This spacing was found sufficient to get converged results, i.e., doubling the number of source points did not lead to visible changes in the flow fields generated. More specifically, the maximum difference between the streamwise velocity fields generated with double the number of source points was less than 3\%.

\begin{figure}
\centering
\includegraphics[trim={0 0 0 0},clip,width=1\linewidth]{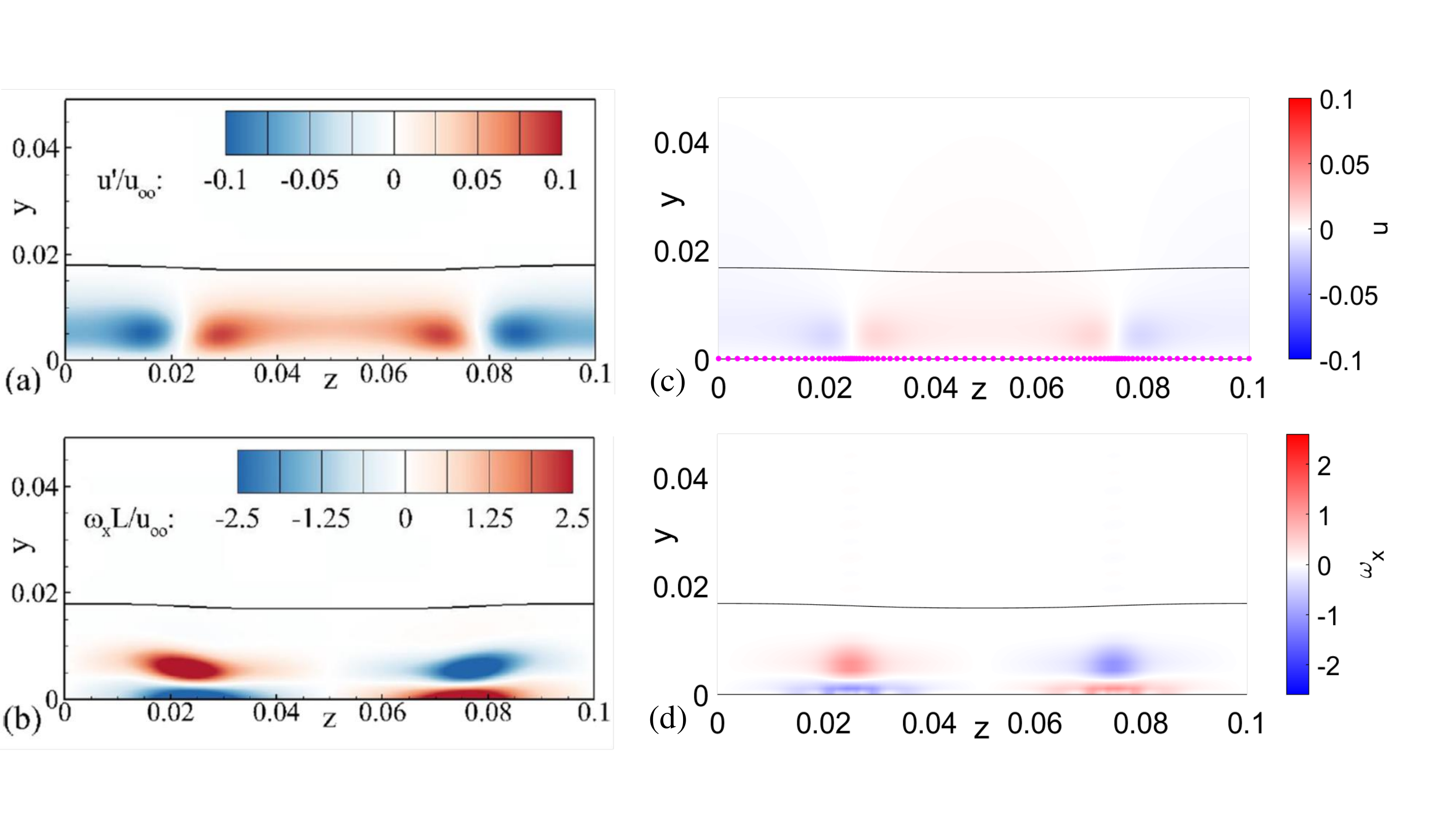}%{Fig_serp_u_vort}
\caption{(a) and (c) Streamwise velocity and (b) and (d) streamwise vorticity fluctuations at $x=1.2$ (here $x,y$ and $z$ are nondimensional values that are normalized by $L=300$~mm). The $99\%$ boundary layer height ($\delta_{99\%}$) is marked by the thick solid line. (a) and (b) Simulation in \citet[figure 10]{RR13}\ftAIP. (c) and (d) Results from the model using the normalizatio from the simulation $u/U=0.1$. The locations of the point sources along the actuation contour are denoted by magenta in (c).}
\label{fig: uvortx_serp}
\end{figure}

Figure~\ref{fig: uvortx_serp} compares the streamwise velocity and vorticity fluctuations obtained through the application of the model in Eq.~\eqref{eq:sum_kxkzeff} with the given forcing functions to the simulation results. These cross-planes of streaks (panel a, and c) and the corresponding streamwise vortical structures (panels b, and d) are shown at a distance $x_0/L=1.2$ from the simulation box inlet (corresponding to a distance of $x=0.2L=\simeq 37\delta^*$ from the actuator's leading edge) obtained through simulations of \citet[figure 10]{RR13} (panels a and b) and from our model (panels c and d). In our model, the output is scaled as $u_{max}/U=0.1$ to match the simulation settings. The plots show qualitative agreement in the shape and location of both the streaks of streamwise velocity and the vortical structures. The high-speed streak location is evident at the spreading point (the front edge of the geometry). The four counter-rotating vortices between the streaks are well captured by the model. However, the magnitude of both, the streak and vortical structures, appears weaker than in the simulations. Also, our model does not capture the slight inclination angle in the horizontal direction that is evident in the two top vortical structures obtained from simulations (panel b). A refinement of the model for this type of actuation, including an extension of the model to account for the 3D-base flows that may address these discrepancies, is left for future work. However, the given results are sufficient to provide an understanding of the important structures that arise due the actuation, which are the key features of the flow manipulation technique.

\section{Concluding Remarks}
 \label{sec:confuture}
 
This paper has provided an extension to the widely used input-output approach that enables its use in evaluating a range of actuation modalities for flow control applications. The advancement lies in exploiting the linearity of the approach to construct flow fields due to localized actuator geometries comprised of a collection of point sources with varying intensity and input directions. The applicable input set was also expanded from the common forcing types (stochastic, harmonic, and impulsive signals) to pulse-width modulated signals that can be adjusted to represent a range of common experimental actuation signals. The model was validated using experimental and DNS data for three different DBD plasma actuator geometries that are commonly used in drag reduction studies. The analytical model was shown to reproduce both the streamwise velocity and vortical structures observed downstream from the actuator in all three configurations. In practice, the goal of creating such structures is to stabilize the flow and delay transition, which in turn can lead to drag reduction. This analysis method can therefore provide insights into the effect of a wide range of flow manipulation strategies that target important structural features such as the streamwise vortices involved in the self-sustaining process. Moreover, these insights can be gained without the expense of detailed experimental or high-fidelity simulations, which are too costly for extensive parametric studies.

 The proposed approach is particularly well suited to control approaches aimed at 
producing the types of streamwise vortical structures commonly used in control aimed at preventing transition and for drag reduction. However, as with all models, there are limitations to the types of problems that this framework is designed to address. The linearization of the dynamics around a base flow makes it less suited to applications aimed at large modifications of the base flow. The approach is also not well suited to control approaches that exploit nonlinear effects. However, some aspects of the nonlinear flow dynamics and different flow regimes can be captured through the inclusion of a modified base profile or an eddy viscosity model \citep{AJ06,MS10,HC10}. Extensions that employ more detailed base flow models or investigate nonlinear interactions between dominant response modes, e.g., using the approach in \cite{DM15,JCDM21}, would allow additional nonlinear effects to be modeled. 

The results here focused on the steady-state response; however, the framework is also directly applicable to other actuation signals (as discussed in Sec.~\ref{sec:iomodel}) and more complex actuation geometries. The input-output framework upon which the proposed approach is based has proven applicable to a wide variety of flow regimes, including those with spatially or temporally periodic base flows as well as both non-Newtonian and compressible flows  (see, e.g., the recent review in \citep{Jo20}). The actuated flow input-output model can be applied to these cases through similar extensions. The technique can also benefit from data-driven turbulence modeling techniques that have already been combined with input-output analysis in a variety of applications (see, e.g., \cite{ZJG17,DIX19}).

Examination of the applicability of this framework in modeling systems with multiple actuators placed at different locations is left for future work.

\begin{acknowledgments}
The authors wish to thank Charles Meneveau for his insightful comments, as well as Thomas Flint and Stanislav Gordeyev for their professional advice regarding plasma actuator operation and modeling. Funding by the US National Science Foundation (CBET 1652244) and the Office of Naval Research (N000141812534) is gratefully acknowledged.
\end{acknowledgments}

\bibliography{IGrefPost}

\end{document}